\documentclass[prb,twocolumn,showpacs,preprintnumbers,amsmath,superscriptaddress,amssymb]{revtex4-1}

\usepackage{graphicx}
\usepackage{epstopdf} 

\usepackage{dcolumn}
\usepackage{bm}
\usepackage[active]{srcltx}
\usepackage{color}

\usepackage{multirow}
\usepackage{amsmath}
\usepackage{amssymb}
\usepackage{amsfonts}
\usepackage{amscd}
\usepackage{multirow}

\begin{document}

\title{Shot noise and conductivity at high bias in bilayer graphene:\\Signatures of electron-optical phonon coupling}

\author{A. Fay}
\affiliation{Low Temperature Laboratory, Aalto University, P.O.Box 15100, FI-00076 AALTO, Finland}
\author{R. Danneau}
\affiliation{Low Temperature Laboratory, Aalto University, P.O.Box 15100, FI-00076 AALTO, Finland}
\affiliation{Institute of Nanotechnology and Institute of Physics,  Karlsruhe Institute of Technology, Germany}
\author{J. K. Viljas}
\affiliation{Low Temperature Laboratory, Aalto University, P.O.Box 15100, FI-00076 AALTO, Finland}
\author{F. Wu}
\altaffiliation{Present address: Department of Microelectronics and Nanoscience, Chalmers University of Technology, SE-41296 G\"oteborg, Sweden.}
\author{M. Y. Tomi}
\author{J. Wengler}
\affiliation{Low Temperature Laboratory, Aalto University, P.O.Box 15100, FI-00076 AALTO, Finland}
\author{M. Wiesner}
\affiliation{Low Temperature Laboratory, Aalto University, P.O.Box 15100, FI-00076 AALTO, Finland}
\affiliation{Faculty of Physics, Adam Mickiewicz University, 61-614 Poznan, Poland}
\author{P. J. Hakonen}
\affiliation{Low Temperature Laboratory, Aalto University, P.O.Box 15100, FI-00076 AALTO, Finland}


\begin{abstract}

We have studied electronic conductivity and shot noise of bilayer
graphene (BLG) sheets at high bias voltages and low bath
  temperature $T_0=4.2$ K. As a function of bias, we find initially
an increase of the differential conductivity, which we
attribute to self-heating. At higher bias, the
conductivity saturates and even decreases due to backscattering from
optical phonons.  The electron-phonon interactions are also
responsible for the decay of the Fano factor at bias voltages
$V>0.1$ V.  The high bias electronic temperature has been
calculated from shot noise measurements, and it goes up to $\sim1200$ K
at $V=0.75$ V. Using the theoretical temperature dependence of BLG
conductivity, we extract an effective electron-optical phonon
scattering time $\tau_{e-op}$. In a 230 nm long BLG sample of mobility
$\mu=3600$ cm$^2$V$^{-1}$s$^{-1}$, we find that $\tau_{e-op}$
decreases with increasing voltage and is close to the charged impurity
scattering time $\tau_{imp}=60$ fs at $V=0.6$ V.
\end{abstract}


\maketitle

\section{Introduction}

Graphene, a two-dimensional plane of carbon atoms arranged in a
honeycomb lattice, has recently attracted wide attention due to its
unique properties.~\cite{CastroNeto2009} High electronic
mobility~\cite{Bolotin2008} up to more than $2\times 10^5$
cm$^2$ V$^{-1}$ s$^{-1}$ combined with a good scalability makes
graphene-based field-effect transistors (FETs) potential basic
building blocks for future nanoelectronics devices. A large on/off
current ratio is required to challenge the current Si-based FETs. In
monolayer graphene (MLG), the on/off current ratio is low ($\sim$5)
because the conductivity always remains above a value of roughly
$4e^2/h$ or $4e^2/\pi h$ for
diffusive and ballistic~\cite{miao2007,Danneau2008} samples,
respectively.
In bilayer graphene (BLG), this ratio can amount up to
20\,000 thanks to the possibility to induce a band gap controlled by
chemical doping \cite{Otah2006} or by a perpendicular electric
field.~\cite{Oostinga2008,Xia2010,Jing2010} The standard FETs commonly work at
high bias voltage where electron-phonon interactions play a major role
in the electronic transport. As a consequence, better knowledge of
the electronic interactions in BLG is of importance for optimizing
graphene-based nanoelectronics devices.

The electronic mobility of BLG samples supported on SiO$_2$ does not
exceed 10\,000 cm$^{2}$V$^{-1}$s$^{-1}$ due to a strong long-range
scattering from charged
impurities.~\cite{Morozov2008,adam2008,Xiao2010} In
suspended~\cite{Feldman2009} or
boron-nitride-supported~\cite{Dean2010,DasSarma2011} BLG samples of
higher mobility, the charged impurities scattering is much weaker and
the short-range scattering (potentially caused by point defects,
neutral scatterers or vacancies) is of importance. We consider here
the low temperature limit ($T_0\sim4$ K) where the electron-phonon
coupling is negligible in the linear response
  regime.~\cite{Adam2010,Viljas2010} With increasing bias voltage,
the BLG differential conductivity in our samples
initially goes up as a result of
self-heating.~\cite{Viljas2011} At high bias voltages,
electron-optical phonon coupling is relevant and considerably influences the electronic transport. In MLG samples,
partial current saturation was reported and related to
electron-optical phonon (e-op) coupling.~\cite{Meric2008,
  Barreiro2009, Perebeinos2010, Bae2010} This coupling was confirmed
by investigating the phonon temperature by Raman
spectroscopy.~\cite{Chae2010, Berciaud2010, Freitag2010} In supported
samples, the population of the different optical phonon modes is
difficult to quantify partly because both the
  intrinsic optical phonons of the graphene and those of the SiO$_2$
  surface can be involved.

In this work, we present electronic conductance and shot noise
measurements on several BLG samples. At high bias voltage, the decrease of both the conductance and the Fano factor are
interpreted as signatures of electron-optical phonon coupling
in BLG. The absence of current saturation is consistent with
  this coupling being weaker compared to that in MLG as calculated by Borysenko \emph{et al.}~\cite{Borysenko2011}
Nevertheless, in this regime we may expect either the electron-phonon
  or electron-electron interactions to be strong enough for establishing a
  quasi-equilibrium electron energy distribution, and consequently an
  estimate of the effective electronic temperature can directly be
extracted from shot noise.~\cite{Santavicca2010,Wu2010} In this way, we find
that the electronic temperature of our BLG samples can go up to
$\sim1200$ K at a voltage of 0.75 V which confirms strong
self-heating~\cite{Viljas2011} and which agrees with optical spectroscopy
experiments.~\cite{Bae2010} Finally, we make use of the electronic
temperature to extract an approximate for the e-op interaction time
($\tau_{e-op}$). We find that this inelastic interaction time
decreases with bias voltage becoming close to the elastic interaction
time (60 fs) at a voltage of 0.6 V in a 230 nm long BLG sample.

\section{Sample fabrication and experimental setup}

The BLG samples have been mechanically exfoliated from graphite by means of a semiconductor wafer dicing tape. A strongly
doped Si substrate, separated by $d=250$ or $270$ nm of SiO$_2$ from
the sample, was used as a back-gate to tune the BLG charge density
$n_g$. The leads were patterned using standard e-beam lithography
techniques and a bilayer resist was employed to facilitate the
lift-off. The samples were contacted using Ti/Al/Ti sandwich
structures with thicknesses 10 nm / $h$ / 5 nm where $h$ was varied
over $50-70$ nm (10 nm of Ti is the contact layer). Metal evaporation
was made in a ultra high vacuum chamber ($10^{-10}$ mBar) in order to obtain the highest contact transparency.~\cite{note_contacts}
Samples of various lengths $L$ from 230 nm up to 1 $\mu \rm{m}$ and widths $W$ from 0.9 to 1.6 $\mu$m were
fabricated. All the data were measured using a two-lead configuration
in a $^4$He dewar at a bath temperature of $T_0=4.2$ K.

Differential AC-conductivity was recorded at frequency $f=32$ Hz with
a standard lock-in technique. The average Fano factor $F \equiv
[S_I(I)- S_I(0)]/2eI$ was calculated by integrating the current
spectral density $S_I$ over the frequency range $f=600-900$ MHz. The
noise generated by the sample was successively amplified
by means of a home-made low-noise amplifier~\cite{Roschier2004}
at 4 K and room-temperature amplifiers by 16 dB and 70 dB,
respectively. The $^4$He dewar was placed in a shielded room in order
to protect noise measurements against external microwave
radiation. After the two amplification stages, the noise was
filtered with a band-pass filter ($600-900$ MHz) and
rectified using a Schottky diode.~\cite{Danneau2008} At low
noise power ($<10$ mW), the DC-voltage measured after the diode is
directly proportional to the noise power.
%
The noise calibration was done by measuring
the shot noise of a tunnel junction which is known to have a Fano
factor $\mathcal{F}=1$ (see Ref.~\onlinecite{blanter2000}). A microwave
switch allowed us to use the same amplification channel for measuring the noise of the graphene
sample and the tunnel junction.~\cite{Danneau2008}

\section{Introduction to shot noise}

In mesoscopic devices, shot noise originates from the granular nature
of charge carriers. Shot noise contains information on the
electronic transport properties that can not be obtained by simple
conductance measurements.~\cite{blanter2000} Moreover, shot noise can be used
to probe the electronic temperature of mesoscopic systems and it is even possible
to use this effect as a primary thermometer.~\cite{spietz03} At the zero frequency
limit, shot noise is given by the correlation function of current
fluctuations $\delta I(t)$: $S_I=\int dt\langle{\delta I(t)\delta
  I(0)}\rangle$. Typically, the strength of shot noise is
characterized by the Fano factor, defined as
$\mathcal{F}=S_I/2e\langle{I}\rangle$. The measured average Fano
factor $F$ is approximately related to the true Fano factor
$\mathcal{F}$ via the Khlus formula:~\cite{Khlus87}
\begin{equation}\label{khlus}
    F \approx \left(\coth \frac{eV}{2k_B T}
    -\frac{2k_B T}{eV} \right) \mathcal{F}.
\end{equation}
Note that when $e|V|\gg k_BT$, the measured $F$ is close to the true
Fano factor ($F\approx\mathcal{F}$). The Fano factor is different
depending on the considered mesoscopic sample. For example, it is $ 1
$ for a low-transparency tunnel junction, $1/3$ for a diffusive metallic
wire~\cite{steinbach1996,henny1999}, $1/4$ for a chaotic
cavity~\cite{Brouwer1996,oberholzer2001,oberholzer2002,silvestrov2003}, and $0$ for a ballistic sample.~\cite{reznikov1995,kumar1996} The shot
noise measurements were used to measure the effective charge of $ e /
3 $ in the fractional quantum Hall regime~\cite{Saminadayar1997,depiciotto1997} and
$2e$ in a SN junction.~\cite{Jehl2000, Kozhevnikov2000} It could also be used to study many-body
interactions and spin related phenomena.~\cite{roche2004,dicarlo2006} Shot noise has
turned out to be very useful in the studies of
graphene.~\cite{Danneau2008,dicarlo2008} For ballistic bilayer
graphene, theoretical calculations give either $\frac{1}{3}$
\cite{snyman2007} or $1-\frac{2}{\pi}$,\cite{katsnelson2006a}
i.e. very close to what has been observed at the charge neutrality point (CNP) in monolayers
\cite{Danneau2008,dicarlo2008} and to the value $\frac{1}{3}$ for
diffusive conductors without inelastic
interactions.~\cite{blanter2000} By reducing the width of MLG sheets down to nanoribbons, a strong suppression of shot noise has been observed consistent with inelastic hopping in these disordered systems.~\cite{danneau2010}

If local temperature $T(x)$ is well defined (quasiequilibrium) at every position $x$ along the length of the graphene sheet, then
it can be shown using theory for
diffusive conductors that (for $e|V|\gg k_BT_0$)
\cite{nagaev1995,naveh1998}
\begin{equation}\label{eq:fano}
\mathcal{F}\approx\frac{2k_BT_{e}}{eV},
\end{equation}
where $T_e\equiv(1/L)\int_0^Ldx T(x)$ is the average temperature.~\cite{footnote1}
The electron-electron (e-e) interactions may lead to an increase of $\mathcal{F}$ over
the value $\frac{1}{3}$.\cite{nagaev1995}
This happens if the power carried away via e-op coupling is small,
such that all the injected power goes to the leads.  Then
$k_BT_{e}\approx
\frac{\sqrt{3}}{8}eV$, \emph{i.e.} {$\mathcal{F}=
\frac{\sqrt{3}}{4}>\frac{1}{3}$,}
the ``hot electron'' value.\cite{blanter2000}
Now, if heat leaks out
from the system also via electron-optical phonon (e-op) coupling, then $T_{e}$ is reduced and hence
$\mathcal{F}$ decreases. Thus, there are two opposite tendencies
for $\mathcal{F}(V)$, but at very large bias the e-op coupling
will dominate and reduce the noise.

\section{Experimental results}
In this paper, we focus on the results obtained on three
different BLG samples. The first one is a 0.23 $\mu$m long and 1.1
$\mu$m wide BLG sample shown in the lower inset of
Fig. \ref{fig:sigma_vs_Vg}. This sample located in between terminals H3 and H4
is named H$34$ in the following. We report also conductance
measurements on two other samples of different length (0.35 $\mu$m and
0.95 $\mu$m) which have been built from the same BLG sheet
(see lower left inset of
Fig.~\ref{fig:sigma_vs_V_diffLength}). Experimental findings reported
on these samples are qualitatively similar to those found on all the
other measured BLG samples.

\subsection{Electronic mobility}

\begin{figure}[htb!]
\begin{center}\includegraphics[width=0.8\linewidth]{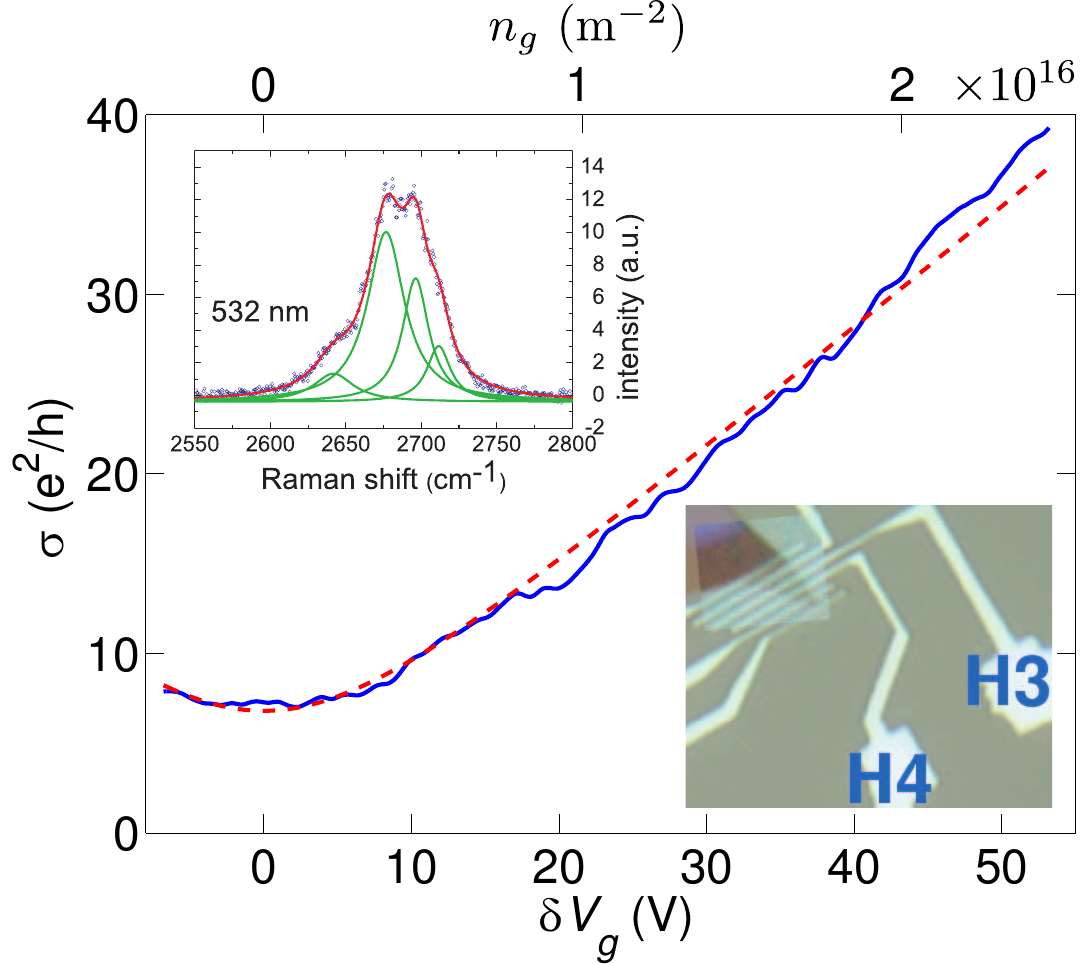}
\caption{(Color online). Conductivity of sample H$34$ as a function of the normalized gate voltage $\delta V_g\equiv V_g-V_{g,min}$, where $V_{g,min}=-23V$ is the value of $V_g$ at the charge neutrality point (CNP). The solid line depicts the experimental data. The dashed line is a fit by using $\sigma=e\mu\sqrt{n_{g}^2+n_{*}^2}$, with the residual density $n_*=4.7\cdot10^{15}\ $m$^{-2}$ and the mobility $\mu=3600$  cm$^{2}$V$^{-1}$s$^{-1}$ (equivalent to an impurity density $n_{imp}=3.5\cdot10^{15}\ $m$^{-2}$). Optical interferogram of sample H$34$ located between leads H3 and H4 (of dimension $0.23\times1.1$ $\mu$m$^2$) is illustrated in the
right inset. To make the sheet more visible, digital contrast enhancement
has been applied. The upper inset displays our typical bilayer Raman
spectrum with four lines fitted.} \label{fig:sigma_vs_Vg}
\end{center}
\end{figure}

In this section, we extract the electronic mobility of sample H$34$
which will be used later in Sec.~\ref{sec:theory} to calculate the
theoretical BLG conductivity. Fig.~\ref{fig:sigma_vs_Vg} presents the
gate voltage dependence of the zero-bias conductivity
  $\sigma=(L/W)(dI/dV)|_{V=0}$ of sample H$34$ at 4.2 K. The
conductivity is minimal and equal to 6.9 $e^2/h$ at the
so-called charge neutrality point at $V_g=V_{g,min}=-23$ V. This
sample is thus $n$-doped in the absence of a gate
  voltage. The gate voltage axis in Fig.~\ref{fig:sigma_vs_Vg} has
been renormalized such that $\delta V_{g}=V_g-V_{g,min}$. In other BLG
samples, we commonly observed an asymmetry between the zero-bias
conductance in the $p$- and $n$-regions, the conductivity being lower
in the $p$-region. This asymmetry is more pronounced in short samples
and probably originates from the $n$-doping by the
leads~\cite{Khomyakov2010} and the ensuing formation of a $p$-$n$
junction close to the contact-graphene interface when BLG is
$p$-doped~\cite{Viljas2011,Huard2008,Barraza-Lopez2010,Khomyakov2010,Lee2008,Mueller2009}
($\delta V_g<0$). The $p$-doped region has not been measured for
sample H$34$ because of a strong initial $n$-doping by impurities and
the leads. The dashed line in Fig.~\ref{fig:sigma_vs_Vg} corresponds to
the zero-bias conductance fit using the (zero-temperature) empirical
relation
$\sigma=e\mu\sqrt{n_{g}^2+n_*^2}$, where $n_*$ is the ``residual``
density and $\mu$ the mobility. In the parallel plate model, the
charge density $n_g$ is related to the gate voltage $\delta V_g$ via
$n_g=C_g\delta V_g/e$, with $C_g=\epsilon_r\epsilon_0/d$ the gate
capacitance per area. In short samples, the screening of the electric
field by the leads decreases the effective gate capacitance. This
modification can be accounted for in the parallel-plate model by
replacing $\epsilon_r=3.9$ by a lower effective dielectric constant
$\epsilon_{eff}$. By solving the Poisson
equation,~\cite{footnote2} we find
$\epsilon_{eff}=0.55\times\epsilon_r=2.1$ and $C_g=85$
aF/$\mu$m$^2$.
%
%
Using this gate capacitance value, we extract the mobility $\mu=3600$
cm$^{2}$V$^{-1}$s$^{-1}$ and the residual density
$n_*=4.7\cdot10^{15}\ $m$^{-2}$. In BLG, the impurity density
$n_{imp}$ is related to the mobility $\mu$ via the relation~\cite{Viljas2011} $n_{imp}=8e/(\pi^2\hbar\mu)$, which gives
$n_{imp}=3.5\cdot10^{15}\ $m$^{-2}$.

\subsection{Bias voltage dependence of BLG conductivity}
\label{sec:exp:conductivity}

\begin{figure}[htp!]
\begin{center}
\includegraphics[width=0.8\linewidth]{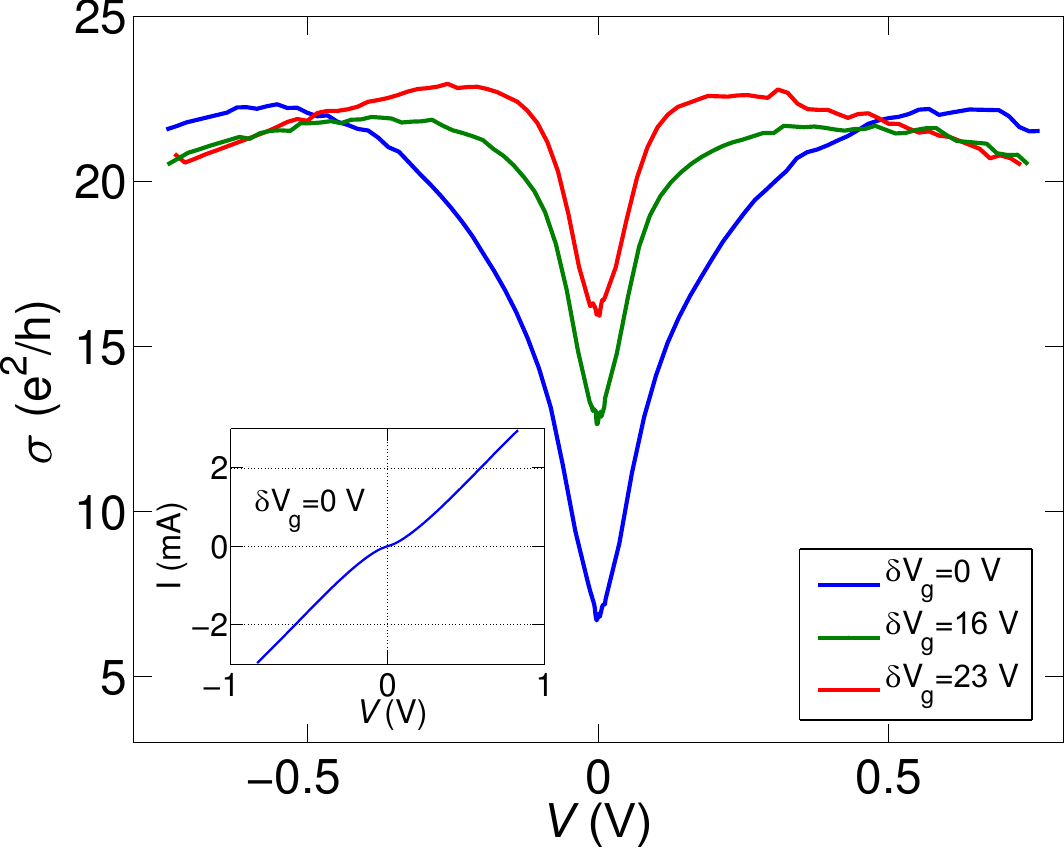}
\caption{(Color online). Conductivity of sample H$34$ as a function of voltage measured at three different gate voltages at $T=4.2$ K. Inset: $I$($V$) curve of sample H$34$ at CNP.}
\label{fig:IV}
\end{center}
\end{figure}

The bias voltage dependence of the differential conductivity
$\sigma(V)=(L/W)(dI/dV)$ of sample H$34$ is shown in Fig.~\ref{fig:IV} for three different gate
voltages. At low bias, the conductivity increases linearly with
voltage, leading to a superlinear current-voltage [$I(V)$]
{characteristic.} This non-linearity is specific to bilayer graphene
and does not show up in monolayer graphene (MLG). It has been
explained in Ref.~\onlinecite{Viljas2011} by {self-heating, though other
contributions cannot} be excluded in the case of this sample. Indeed,
the conductivity in BLG is more sensitive to temperature than in MLG
due to a higher density of states.~\cite{adam2008} Above 0.1 V, the
BLG conductivity grows {slower and slower as the voltage increases,
and reaches a maximum at $0.2$ V$\ldots 0.75$ V.
Above this voltage the conductivity begins to slowly decrease.}

\begin{figure}[htp!]
\begin{center}
\includegraphics[width=0.8\linewidth]{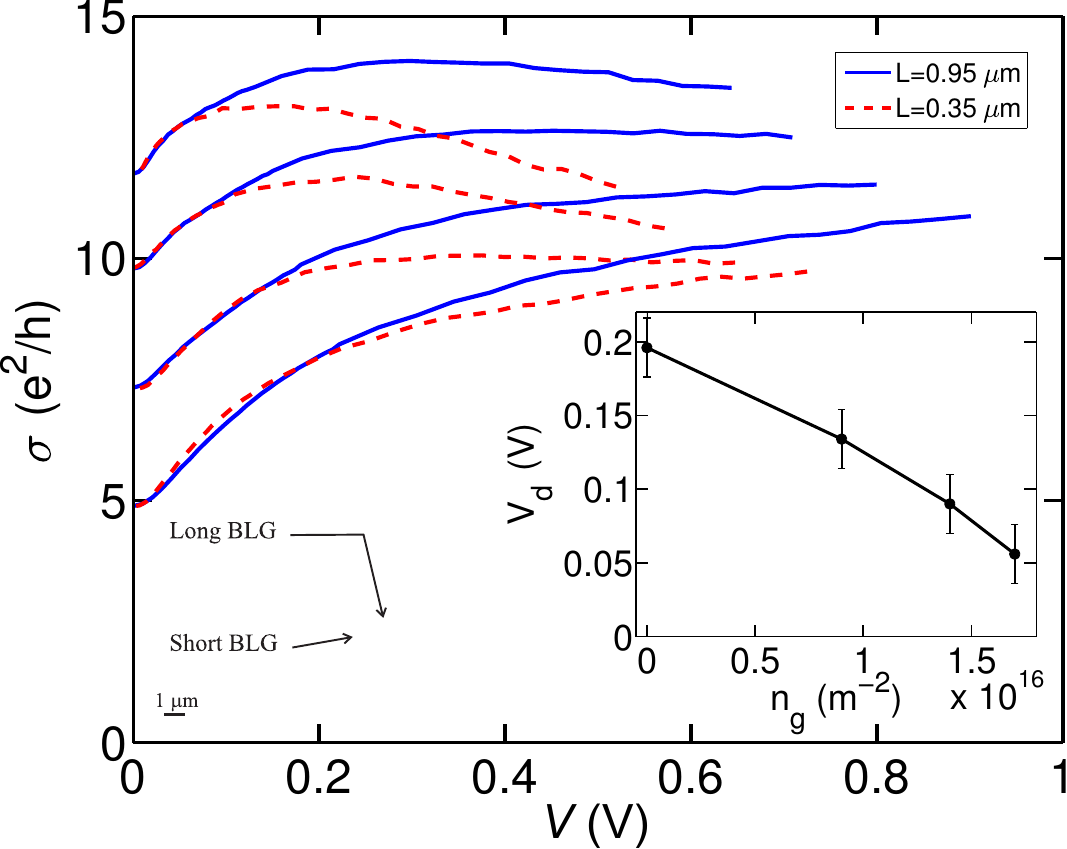}
\caption{(Color online). Conductivity vs. voltage for two BLG samples of different lengths (350 and 950 nm) at 4 different charge densities (from bottom to top curve pairs: $n_g=0,\ 0.9,\ 1.4\ \rm{and}\ 1.75\ \times10^{16}$ m$^{-2}$). Left inset: optical image of the short and long samples made from the same BLG flake, which is colored in red to enhance its visibility. Right inset: voltage $V_d$ at which the conductivities of the short and long samples start to deviate.}
\label{fig:sigma_vs_V_diffLength}
\end{center}
\end{figure}

We also investigated the transport properties of 2 BLG samples of
different length (350 nm and 950 nm) made from the same BLG
sheet.~\cite{Viljas2011} As shown in
Fig.~\ref{fig:sigma_vs_V_diffLength}, at low bias and for the same
charge density, the conductivity does not depend on the length $L$ of
the sample, showing the same initial increase with voltage. However at
higher bias, the conductivity of the short sample reaches a maximum
and then decreases faster than in the long sample. We define $V_d$ as the
voltage above which the conductivities of the short and long
samples deviate.

In order to explain these experimental results, we now focus on the
backscattering induced by interactions between electrons and optical
phonons (op). Note that extrinsic surface phonons modes of the
SiO$_{2}$ substrate are here relevant and may be the primary source of energy relaxation.~\cite{Chen2008,Fratini2008}. In a simple model, we assume
that the electrons are accelerated by the external electric field
$\mathcal{E}\equiv V/L$, so the energy gained by the electron after a
distance $x$ is $e|\mathcal{E}|x$. When this energy reaches the characteristic optical phonon energy $\hbar\Omega$, electrons can transmit their
energy to optical phonons (op). Assuming an instantaneous energy exchange, the
``threshold'' mean-free-path for the onset of scattering
is thus given by~\cite{Yao2000}
$x\sim \hbar\Omega/e\mathcal{E}$.
At the same
bias voltage, the electric field is higher in the short sample and,
consequently, the threshold mean free path is shorter. This explains why
the conductivity of the short sample is lower than that of the long
sample. At low bias voltages, the fact that the conductivities of both
samples are almost identical is in agreement with a self-heating
effect~\cite{Viljas2011} in absence of electron-phonon
coupling. The voltage $V_d$ is lowered when the charge density
increases [inset of Fig.~\ref{fig:sigma_vs_V_diffLength}], suggesting an
enhancement of backscattering from phonons far from the CNP.

\subsection{Shot noise in BLG}

\begin{figure}[htp!]
\begin{center}
\includegraphics[width=0.85\linewidth]{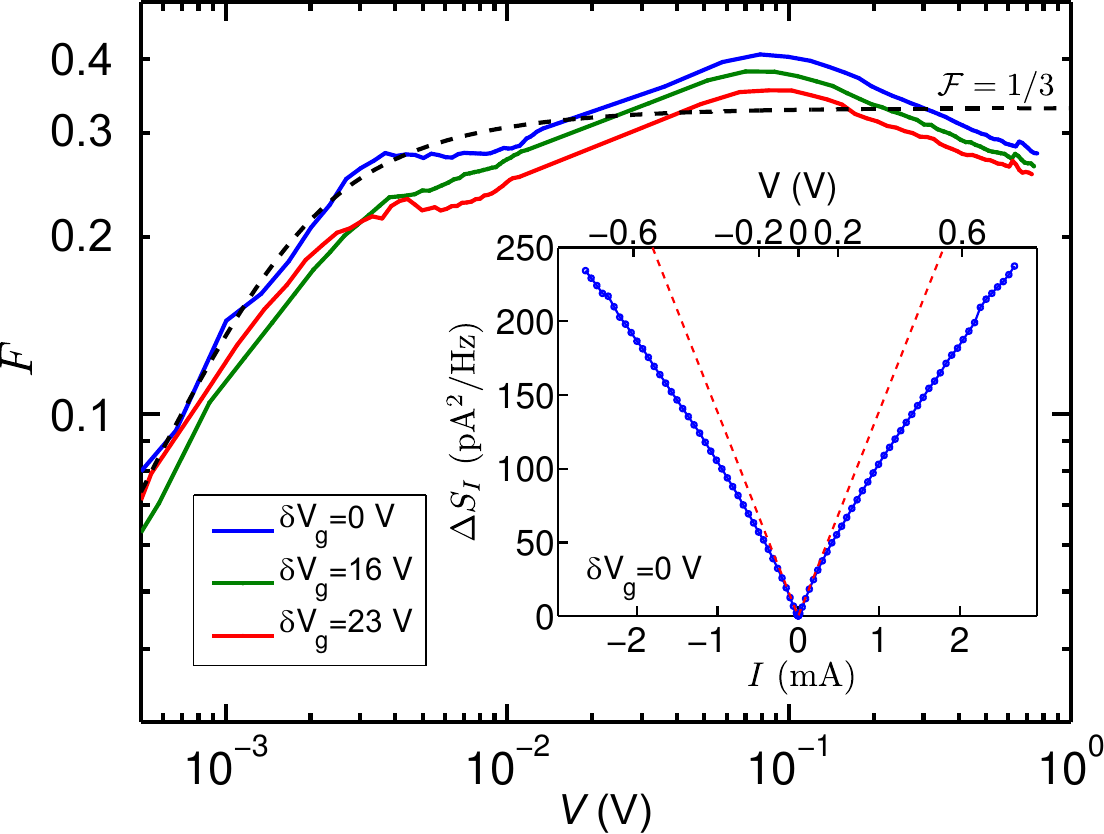}
\caption{(Color online) Excess noise Fano factor $F = [S_I(I)- S_I(0)]/(2eI)$
vs. $V$ at three values of $\delta V_{g}$ indicated in the
figure. The dashed curve is calculated using Eq.\ (\ref{khlus}) for $\mathcal{F}=1/3$.
Inset: $\Delta S_I=S_I(I)-S_I(0)$ as a function of the current $I$ at CNP. The dashed lines depict the theoretical values of $\Delta S_I(I)=\mathcal{F}2eI$ with a constant Fano factor $\mathcal{F}=\sqrt{3}/4$, making clear the drop of $\mathcal{F}$ above $V\sim 0.2$ V. The non-linearity of the voltage axis (upper axis) comes from the non-linear $I(V)$ curve.}
\label{fig:F_vs_V}
\end{center}
\end{figure}

Additional information to electronic conductivity measurements was obtained from shot noise in BLG. In particular, an
  effective electronic temperature was directly obtained from
{shot noise}, which is analyzed in Sec.~\ref{sec:teop} to extract
{an estimate for the} e-op interaction time. Fig.~\ref{fig:F_vs_V}
presents the voltage dependence of the average Fano factor $F$ for
the same gate voltages as for the conductivity curves in
Fig.~\ref{fig:IV}. At low bias voltages $V\lesssim2$ mV, $F$ increases
linearly with voltage as expected from the Khlus formula \eqref{khlus}
when $eV\ll2k_BT$: $F\approx\mathcal{F}eV/6k_BT$. By fitting the noise at low bias with the Khlus formula with a fixed temperature $T=4.2$ K,
we find that $\mathcal{F}$ is close to $1/3$ at the CNP as expected for a diffusive sample,~\cite{blanter2000} and it is slightly lower far from the CNP which may be related to disorder~\cite{SanJose2007,Tworzydlo2008,Lewenkopf2008,Titov2010} ($\mathcal{F}\sim0.3$ at $\delta V_g=16$ and $23$ V.). At bias
voltages above 20 mV, the average Fano factor is weakly dependent on gate voltage, and $F$ first goes higher than predicted from the Khlus formula. We interpret this as a signature of
a hot-electron regime.~\cite{steinbach1996} Above 0.1 V, $F$ decays
and goes to a value of $\sim0.26$ at $V=0.75$ V. We
assign the decrease of the Fano factor at high bias voltages to
the coupling between electrons and optical phonons.

\section{Electron-optical phonon scattering time}
\label{sec:theory}

The theoretical conductivity for BLG is a function of the chemical
potential $E_F$, the average electronic temperature $T_e$ and the
electron scattering time $\tau$, which is independent of energy for
screened Coulomb impurities.~\cite{Viljas2011}
Introducing again the phenomenological residual density $n_*$, we
may write the conductivity as
\begin{equation}
\sigma(\tau,E_F,T_e)=e\mu\sqrt{(n_e+n_h)^2+n_*^2},
\label{eq:conductivity}
\end{equation}
with $n_e=\{\gamma_1k_BT_e/[\pi(\hbar v_0)^2]\}\ln(1+e^{E_F/k_BT_e})$
and $n_h=\{\gamma_1k_BT_e/[\pi(\hbar v_0)^2]\}\ln(1+e^{-E_F/k_BT_e})$
the electron and hole concentrations, respectively.  The chemical
potential $E_F$ is related to the charge density $n_g=n_e-n_h$ via the
relation $E_F=[\pi(\hbar v_0)^2/\gamma_1]n_g$. We assume that electron
and hole mobilities are identical~\cite{Viljas2011} and
given by $\mu=(2e{v_0}^2/\gamma_1)\tau$. Using Matthiessen's rule, we
simply replace the scattering rate $\tau^{-1}$ by the sum of two
contributions: $\tau^{-1}=\tau_{imp}^{-1}+\tau_{e-op}^{-1}$, coming
from charged impurity and electron-phonon scattering, respectively. At
zero bias, charged impurity scattering
dominates,~\cite{Adam2010,Viljas2010} so $\tau(0
\ \rm{V})=\tau_{imp}$. Using the zero bias mobility $\mu=3600$
cm$^2$V$^{-1}$s$^{-1}$, we deduce $\tau(0 \ \rm{V})=60$ fs. We
assume next that $\tau_{imp}$ does not depend on voltage.

In the following, we first consider only scattering from charged
impurities, neglecting inelastic processes, i.e. $\tau=\tau_{imp}$. In
order to calculate the conductivity from Eq.~\eqref{eq:conductivity},
the electronic temperature ($T_e$) has to be known. In the next
section, $T_e$ is extracted at low bias by solving the heat diffusion equation in absence of electron-phonon coupling and at high bias from shot noise measurements (in presence of e-op coupling).

\subsection{Electronic temperature}
\begin{figure}[htp!]
\begin{center}
\includegraphics[width=0.8\linewidth]{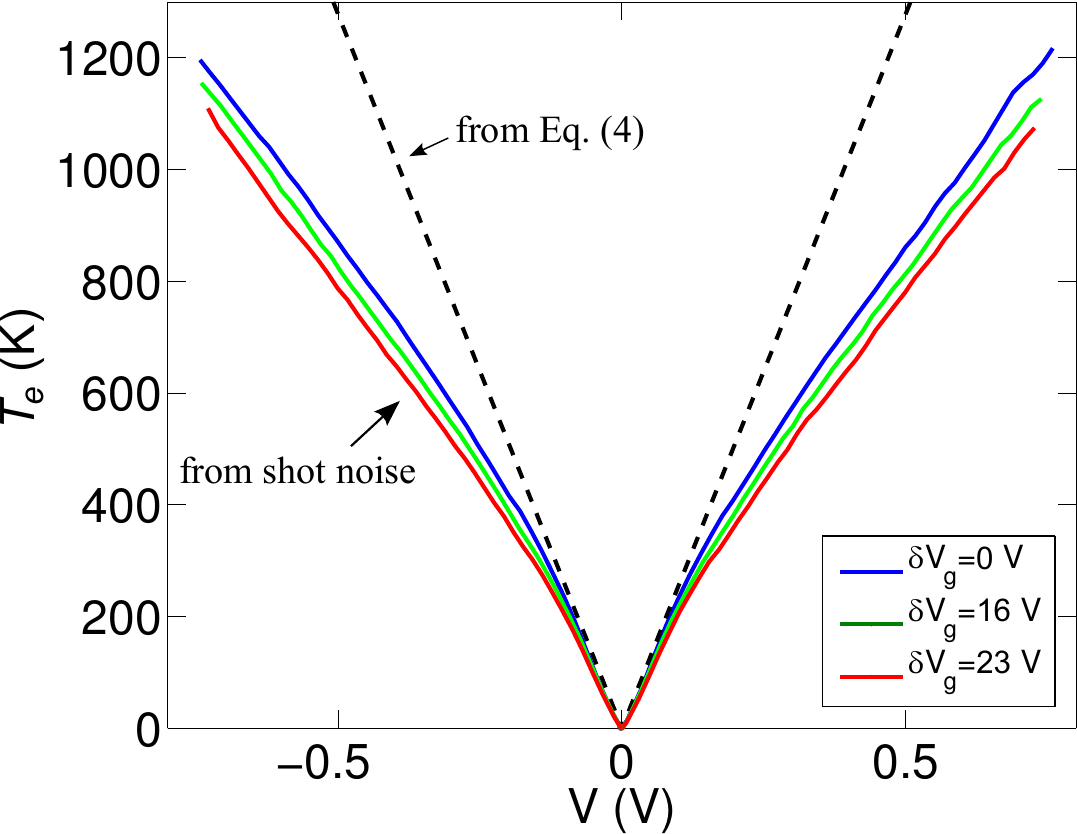}
\caption{(Color online). The solid lines are the electronic temperature extracted from the Fano factor [$T_e=eV\mathcal{F}/(2k_B)$] as a function of voltage at three different gate voltages given in the inset. The dashed line depicts the gate voltage independent average temperature calculated using Eq.~\eqref{eq:temperature_without_eph} in absence of e-op coupling.} \label{fig:T_vs_V}
\end{center}
\end{figure}

As shown in Sec.~\ref{sec:exp:conductivity}, at low
bias voltage the electron-phonon interactions can be neglected. Under
this condition, assuming that the electron energy distribution
function is at quasi-equilibrium, the electronic temperature at the
position $x$ along the BLG sheet is given by solving the heat diffusion equation, using
the Wiedemann-Franz law. In the limit where $T_e\gg T_0=4$ K, it yields
\begin{equation}
T(x)=\sqrt{(x/L)(1-x/L)/\mathcal{L}}\,V,
\label{eq:temperature_without_eph}
\end{equation}
where $\mathcal{L}=\pi^2k_B^2/(3e^2)$ is the Lorenz number.
The average temperature $T_{e}=\frac{\sqrt{3}}{8}eV/k_B$
increases linearly with $V$ and it goes up to 250 K at $0.1$ V [see
  dashed line in Fig.~\ref{fig:T_vs_V}].

Above $\sim0.1$ V, electrons interact with optical phonons, and thus
the average temperature calculated from Eq.~\eqref{eq:temperature_without_eph} is an overestimate. In
this inelastic interaction regime, the electronic temperature
can be extracted from the Fano factor using Eq.~\eqref{eq:fano}. We
find $T_e\sim1200$ K at $0.75$ V which is much lower than the temperature
of 2300 K we would observe without energy relaxation to phonons
[dashed line in Fig.~\ref{fig:T_vs_V}]. Note that the electronic
temperature does not depend much on gate voltage at high bias. This
can be understood from the fact that the broadening of the electron
energy distribution ($\sim k_BT$) is much larger than the chemical
potential~$E_F$.

\subsection{Theoretical conductivity and electron-optical phonon scattering time}
\label{sec:teop}

\begin{figure}[htp!]
\begin{center}\includegraphics[width=0.85\linewidth]{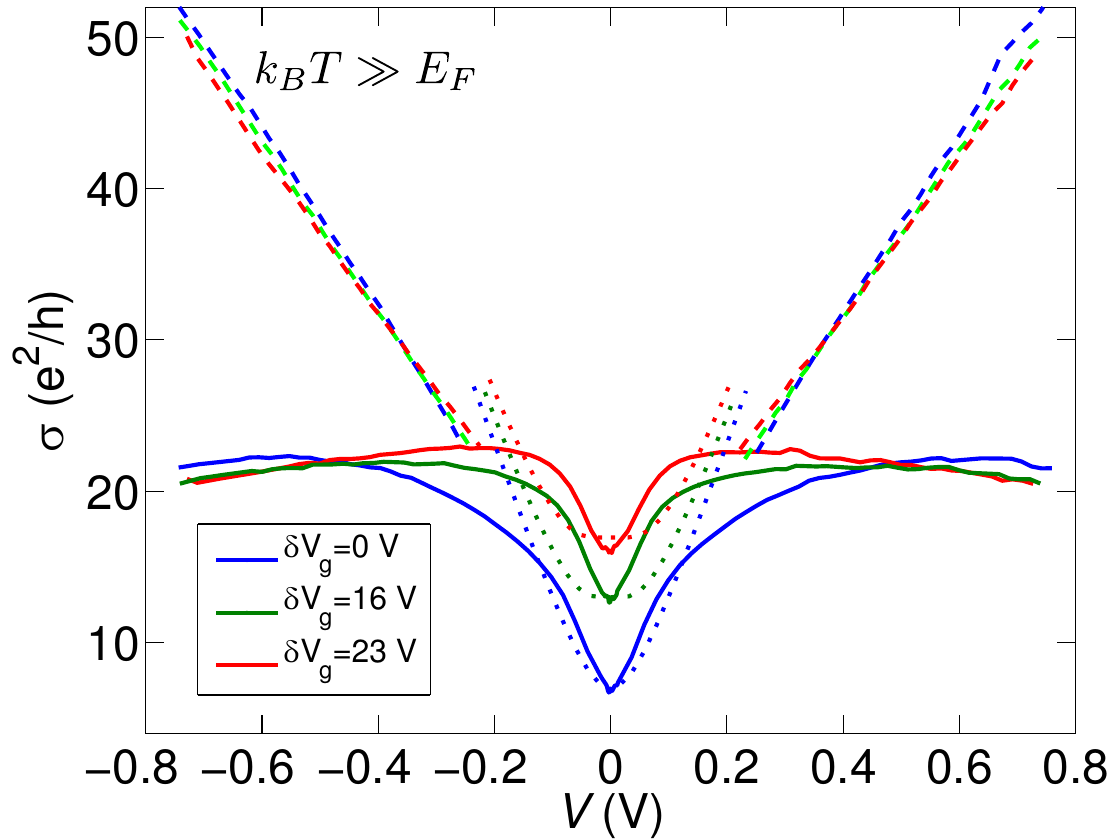}
\caption{(Color online). The solid lines depict the measured conductivity as a function of voltage at three different gate voltages. The dotted and dashed lines are the theoretical conductivities (Eq.~\eqref{eq:conductivity}) calculated using the temperature obtained from Eq.~\eqref{eq:temperature_without_eph} and from the Fano factor, respectively.} \label{fig:sigma_vs_V_theory}.
\end{center}
\end{figure}

The dotted lines shown in Fig.~\ref{fig:sigma_vs_V_theory} depict the
theoretical conductivities at low bias as a function of voltage
calculated without any free parameters by using
Eq.~\eqref{eq:conductivity} and the average temperature $T_e$ from Eq.~\eqref{eq:temperature_without_eph}.
The theoretical
conductivity does not change much at low bias and increases at higher
bias linearly with voltage. This leads to a {low bias
  conductivity ``plateau''} whose width grows with increasing the
chemical potential $E_F$. This conductivity plateau has not been
observed in any of our measured BLG samples as discussed in Ref.~\onlinecite{Viljas2011}.

At high bias voltages ($V>0.2$ V), by using the temperature
extracted from the Fano factor, we find that the theoretical
conductivity given by Eq.~(\ref{eq:conductivity}) increases linearly with
increasing $V$ and is almost always above the experimental
conductivity (dashed lines in Fig.~\ref{fig:sigma_vs_V_theory}). As a
result, the discrepancy between the theoretical and experimental
conductivities increases with increasing $V$. This was to be expected as
we have only considered so far charged impurity scattering and ignored
e-op phonon interactions.

\begin{figure}[htp!]
\begin{center}
\includegraphics[width=0.85\linewidth]{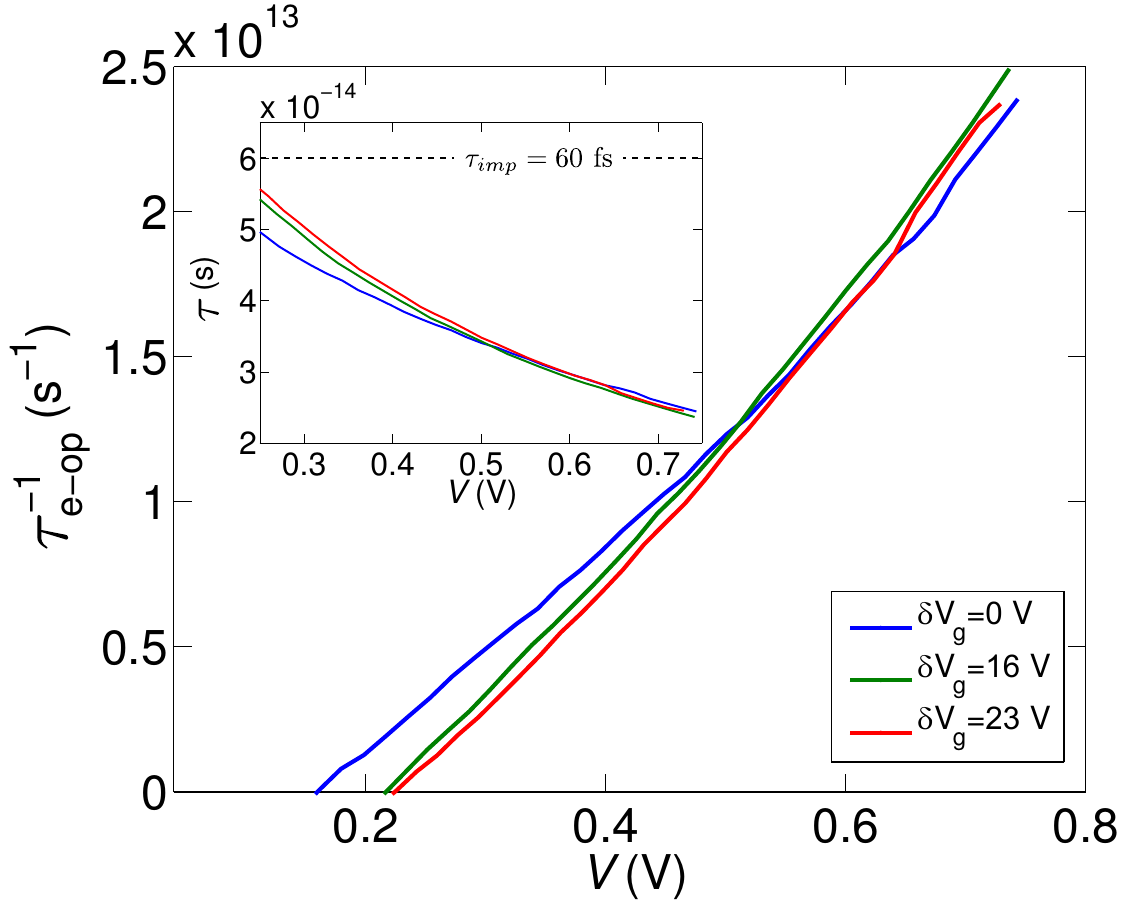}
\caption{(Color online). Electron-phonon scattering rate $\tau_{e-op}^{-1}$ as a function of voltage in sample H$34$. Inset: voltage dependence of the total scattering time $\tau=(\tau_{imp}^{-1}+\tau_{e-op}^{-1})^{-1}$.} \label{fig:tau_vs_V}
\end{center}
\end{figure}

We next fit the experimental conductivity with
Eq.~\eqref{eq:conductivity} by adjusting the interaction time
$\tau_{e-op}$. Note that the temperature extracted from the Fano
factor takes into account e-op interactions and, thus, is still
valid. At $V\sim0.2$~V, the theoretical and experimental
conductivities are close, resulting in a large uncertainty for
$\tau_{e-op}$. We find that the scattering rate $\tau_{e-op}^{-1}$ does not depend much on gate
voltage (Fig.~\ref{fig:tau_vs_V}). This can again be understood from the
fact that $k_BT\gg E_F$. The rate $\tau_{e-op}^{-1}$ is very low ($<0.2\times10^{13}$ s$^{-1}$) at voltages below $0.2$ V and charged impurity scattering dominates. Above $V=0.2$ V, $\tau_{e-op}^{-1}$ increases almost linearly
with increasing the bias voltage, as calculated for MLG.~\cite{tse2008} At $V=0.75$ V, the inelastic rate is $\sim1/40$ fs$^{-1}$ which is close to $\tau_{imp}^{-1}=1/60$ fs$^{-1}$. The increase of the inelastic scattering rate $\tau_{e-op}^{-1}$ can be explained by the presence of more phase space for the e-op scattering.

\section{Conclusion}

In conclusion, we have investigated conductance and {shot
  noise} in several BLG samples at a bath temperature of 4.2~K. At low
bias voltage, the increase of the {differential} conductivity
with voltage is attributed to self-heating. In this
regime, the scattering time is limited by charged impurity
scattering. At high bias voltage, decrease in the
conductance indicates electron-optical phonon
{scattering} which is also confirmed by reduction of shot noise. The drop of the conductivity is more pronounced
in short samples suggesting a larger e-op scattering rate. This agrees
with the {expectation that the threshold mean-free-path} for e-op
coupling is directly proportional to
the sample length. At high bias voltage, the electronic temperature
$T_e$ in the BLG sheet is deduced from shot noise measurements. The
electronic temperature goes up to $\sim$1200 K at a voltage of $0.75$
V. From the temperature $T_e$, we {estimated} the e-op scattering time
and found that it decreases with increasing voltage. In a 230 nm long
BLG sample, we find that $\tau_{e-op}$ goes below the elastic
scattering time $\tau_{imp}=60$ fs at a voltage of $0.6$ V. The e-op
coupling is thus manifested at the standard working points of
graphene-based FETs. Extended work on suspended graphene samples is of
interest in order to suppress the relaxation channel to SiO$_2$
surface phonons.

\section*{Aknowledgment}
This work was supported by the Academy of Finland, the European Science Foundation (ESF) under EUROCORES Programme EuroGraphene and
the EU project MMM@HPC FP7-261594. R.D.'s Shared Research Group SRG 1-33 received financial support by the Karlsruhe Institute of Technology within the framework of the German Excellence Initiative.


\begin{thebibliography}
\bibitem{note_contacts} We have measured maximum contact resistances on the order of 20 $\Omega\mu \rm{m}^2$.
\bibitem{footnote1} {Using Eq.\ (\ref{eq:fano}) for graphene close to the CNP is an
approximation, because the derivation of the result requires a constant density of states.}
\end{thebibliography}
\end{document}